\documentclass[prd,aps, reprint,nofootinbib]{revtex4-1}
\usepackage{geometry}                % See geometry.pdf to learn the layout options. There are lots.
\usepackage{graphicx}
\usepackage{amssymb}
\usepackage{mathtools}
\usepackage{epstopdf}
\usepackage{verbatim}
\usepackage{color}
\usepackage{slashed}
\usepackage{bbold}
\usepackage{fullpage}
\usepackage{float}
\usepackage[title]{appendix}
\usepackage[colorlinks, citecolor=red, linkcolor=blue]{hyperref}

\begin{document}

\begin{flushright}\small{MCTP-15-11}\end{flushright}
\title{\textbf{\large{Reaching for Squarks and Gauginos at a 100 TeV p-p Collider}}}

%\author{Sebastian~A.~R.~Ellis\footnote{sarellis@umich.edu}}

%\author{Bob Zheng\footnote{byzheng@umich.edu}}

\author{Sebastian~A.~R.~Ellis}
\email{sarellis@umich.edu}

\author{Bob Zheng}
\email{byzheng@umich.edu}

%\affil{\it{Michigan Center for Theoretical Physics (MCTP),}\\ \it{Department of Physics, University of Michigan}, \\ \it{Ann Arbor, MI 48109, USA}}

\affiliation{\it{Michigan Center for Theoretical Physics (MCTP),}\\ \it{Department of Physics, University of Michigan}, \\ \it{Ann Arbor, MI 48109, USA}}

\begin{abstract}
We analyse the prospect of extending the reach for squarks and gauginos via associated production at a $\sqrt{s} = 100$ TeV proton-proton collider, given 3 ab$^{-1}$ integrated luminosity. Depending on the gluino mass, the discovery reach for squarks in associated production with a gluino can be up to 37 TeV for compressed spectra (small gluino-LSP mass splitting), and up to 32 TeV for non-compressed spectra. The discovery reach for Winos can be up to between 3.5 and 6 TeV depending on squark masses and Wino decay kinematics. Binos of up to 1.7 TeV could similarly be discovered. Squark-gaugino associated production could prove to be the discovery mode for supersymmetry at a 100 TeV collider in a large region of parameter space. 
\end{abstract}

\maketitle

\section{Introduction}

Observational evidence for low energy Supersymmetry (SUSY) remains elusive. Current LHC data constrains strongly-interacting superpartner masses to lie near or above a TeV, disfavoring electroweak-scale SUSY in a wide variety of models. It is therefore becoming increasingly well motivated to consider the possibility that the superpartner masses lie above $\sim 1$ TeV, perhaps evading the kinematic reach of LHC-14. This has prompted numerous studies of the SUSY discovery potential of future hadron colliders, which have demonstrated that a $\sqrt{s} = 100$ TeV collider can extend the kinematic reach for superpartners into the multi-TeV range \cite{Cohen:2013xda, Jung:2013zya, Low:2014cba, Cohen:2014hxa, Ellis:2014kla, Acharya:2014pua, Gori:2014oua, Bramante:2014tba, diCortona:2014yua, Berlin:2015aba, Beauchesne:2015jra}.

Previous studies of SUSY at future hadron colliders have focused primarily on pair production, either of colored superpartners \cite{Cohen:2013xda, Jung:2013zya, Cohen:2014hxa} or of electroweak-inos \cite{Low:2014cba, Acharya:2014pua, Gori:2014oua, Bramante:2014tba, diCortona:2014yua, Berlin:2015aba}. In this paper, we instead examine the reach of a $\sqrt{s} = 100$ TeV collider for associated production of a heavy squark along with a lighter gaugino. This production channel is particularly noteworthy if the squark masses are $\mathcal{O}(10)$'s of TeV, such that squark pair production is kinematically inaccessible at $\sqrt{s} = 100$ TeV. Spectra where squarks are hierarchically heavier than the gluino/electroweak-inos are predicted in many SUSY breaking models such as anomaly mediation \cite{Randall:1998uk, Giudice:1998xp} or more general ``mini-split"-type scenarios \cite{Wells:2003tf, Acharya:2007rc, Arvanitaki:2012ps}. Moreover, multi-TeV squark masses can naturally accommodate the stop masses required to achieve a 125 GeV Higgs boson within the MSSM.

In this paper, we demonstrate that associated squark-gaugino production at a $\sqrt{s} = 100$ TeV proton collider provides a probe of $\gtrsim 10$ TeV squark masses which is complementary to pair production. Our main results are summarized in Figures \ref{NonComp.FIG}-\ref{SqWinoNLSP.FIG}, which show the reach of a $\sqrt{s} = 100$ TeV p-p collider with 3 ab$^{-1}$ integrated luminosity for squark-gaugino associated production in various spectra \footnote{Note that a recent study in \cite{Hinchliffe:2015qma} calls for an integrated luminosity of between 10 and 20 ab$^{-1}$ at a future 100 TeV p-p collider. We present here results for 3 ab$^{-1}$ as a conservative estimate, and so as to be directly comparable with the current literature.}. 

Squark-gluino production can discover squark masses up to 32 TeV for $\lesssim 4$ TeV gluino masses in spectra with a large gluino-neutralino LSP mass splitting (Fig \ref{NonComp.FIG}). For spectra with a small gluino-neutralino LSP mass splitting, squark masses up to 37 TeV can similarly be discovered (Fig. \ref{Comp.FIG}). Notably, our analysis finds that the gluino-neutralino DM coannihilation region \cite{Profumo:2004wk, Ellis:2015vaa} can be excluded for squark masses $\lesssim 28$ TeV. For squark-Wino (Bino) LSP production, Wino (Bino) masses up to $4\, (1.7)$ TeV  can be discovered for squark masses $\lesssim 7\, (5)$ TeV (Figs. \ref{SqWino.FIG}-\ref{SqBino.FIG}). We find a similar reach for squark-Wino NLSP production  (Fig \ref{SqWinoNLSP.FIG}), even without utilizing objects resulting from NLSP $\rightarrow$ LSP decay. Our results indicate that squark-gaugino production represents a SUSY discovery mode at a $\sqrt{s} =$ 100 TeV p-p collider in a wide variety of models with heavy first- and second-generation squarks. 

The remainder of this paper is organized as follows. Section \ref{general} discusses our general methodology and simulation strategies. Section \ref{sqgo} presents in detail our analysis of squark-gluino associated production, while Section \ref{sqew} presents our analysis of  squark-Wino/Bino associated production. We summarize our results in Section \ref{conclusion}.

\section{General Methodology}\label{general}

In this section we briefly discuss the general methodology of the analyses presented below. Event topologies arising from heavy squark - light gaugino associated production are characterized by a hard leading jet and significant $\slashed{\it{E}}_T$. These objects result primarily from the squark decay products, as the associated gaugino is produced at relatively low transverse momentum. The dominant SM background for such events is in the $t \overline{t} + $ jets and vector boson + jets channels \cite{Cohen:2013xda}, which fall off rapidly with increasing leading jet $p_T$, $\slashed{\it{E}}_T$, and $\slashed{\it{E}}_T/\sqrt{H_T}$ ($H_T$ is defined as the scalar sum of the jet transverse energies).  

In the following analyses, we consider the reach of a $\sqrt{s} = 100$ TeV proton-proton collider given 3 ab$^{-1}$ integrated luminosity. The minimum production cross section yielding $\gtrsim 10$ events is roughly $ \sim 10^{-2}$ fb, corresponding to $m_{\tilde{q}} + m_{\tilde{g}} \sim 35$ TeV  ($m_{\tilde{q}} + m_{\tilde{W}} \sim 15$ TeV) for squark-gluino (squark-Wino) associated production. For such masses, good background discrimination is achieved with hard leading jet $p_T$ cuts for squark-gluino production, and with hard $\slashed{\it{E}}_T/\sqrt{H_T}$ cuts for squark-Wino/Bino production. Our strategy is as follows: for each analysis we impose a set of baseline cuts catered to a set of spectra. We then scan over leading jet $p_T$ and $\slashed{\it{E}}_T$ cuts (squark-gluino) or $\slashed{\it{E}}_T/\sqrt{H_T}$ cuts (squark-Wino/Bino) to maximize significance $\sigma$, defined by\begin{equation}\label{sig}
\sigma \equiv \frac{S}{\sqrt{1 + B + \lambda^2 B^2 + \gamma^2 S^2}}\,\,.
\end{equation} $S$ $(B)$ is the number of signal (background) events passing cuts, and $\gamma$ $(\lambda)$ parameterize systematic uncertainties associated with signal (background) normalization. Details of the event generation and collider simulation are given in Appendix \ref{GenDetails.APP}. Like most future collider studies, our simulated $\sigma$ values are subject to $\mathcal{O}(1)$ uncertainties associated with e.g. the performance of a detector which is yet to be designed. However, this translates to a comparatively mild uncertainty for the predicted reach, due to the rapid falling of production cross sections with increasing mass. 

\subsection*{Simplified Models}

In the analyses presented below, we consider the following SUSY simplified models:
\begin{table}[H]
\centering
\begin{tabular}{ | c | c | c|}
\hline
\textbf{Model} & \textbf{Particle Content} & \textbf{Fig.}\\
\hline
\hline
{\bf Squark-Gluino} & $\tilde{q}$, $\tilde{g}$, $\chi_1^0 = \tilde{B}$ &   \\
Non-compressed & $M_1 = 100$ GeV  & Fig. \ref{NonComp.FIG} \\
Compressed & $m_{\tilde{g}} - m_{\chi^0_1} =15$ GeV  & Fig. \ref{Comp.FIG} \\
\hline
{\bf Squark-Wino LSP}& $\tilde{q}$, $\chi_1^0 = \tilde{W}$ &  Fig. \ref{SqWino.FIG} \\
\hline
{\bf Squark-Bino LSP}& $\tilde{q}$, $\chi_1^0 = \tilde{B}$ & Fig. \ref{SqBino.FIG} \\
\hline
{\bf Squark-Wino NLSP}& $\tilde{q}$, NLSP = $\tilde{W}$, $\chi^0_1= \tilde{B}/\tilde{H}$ &  Fig. \ref{SqWinoNLSP.FIG}\\
Split & $M_1 / \mu = 100$ GeV  & \\
Non-split & $m_{\tilde{W}}-m_{\chi^0_1}= 200$ GeV  & \\
\hline
\end{tabular}
\caption{Simplified models considered in this paper.}
\label{SimpModels.TAB}
\end{table}
\noindent which encompass a wide array of potential event topologies arising from squark-gaugino production. We take degenerate first and second generation squark masses, and decouple all sparticles not listed in Table \ref{SimpModels.TAB}. For the squark-gluino non-compressed model, our results are not sensitive to the choice of $M_1 = 100$ GeV as the LSP is effectively massless for $m_{\chi^0_1} \ll m_{\tilde{g}}$. The squark-gluino compressed model is motivated by the gluino-neutralino coannihilation region \cite{Profumo:2004wk, Ellis:2015vaa}. We choose $ m_{\tilde{g}} - m_{\chi^0_1}= 15$ GeV as a fiducial value, though the leading jet $p_T$-based analysis presented below is robust as long as $m_{\tilde{g}} - m_{\chi^0_1} \ll m_{\tilde{g}}$. For the Wino NLSP models, we choose two spectra with differing LSP masses to illustrate the effects of increasing the NLSP-LSP mass splitting. In the ``non-split" case, we have chosen an NLSP-LSP mass splitting of 200 GeV so that the NLSP decays to the LSP + on-shell SM bosons.

\begin{comment} While such models do not necessarily reflect the full range of phenomenological features possible in other more complicated SUSY constructions, they provide good benchmarks which capture the main qualitative features of more complicated event topologies.

We fix the Bino mass in the non-compressed squark-gluino case to be 100 GeV so that it is much lighter than the other sparticles in the spectrum. Loop effects result in an LSP mass of $\sim 80$ GeV. In the compressed squark-gluino case we select a $m_{\tilde{g}}-m_{LSP} = 15$ GeV mass difference motivated both by consideration of the gluino-Bino coannihilation strip. For the squark-Wino/Bino models, we consider spectra with varying Wino/Bino mass as well as varying squark mass. In the Wino NLSP models, we fix the Bino/Higgsino mass in the split cases to be 100 GeV so that it is much lighter than the other sparticles in the spectrum, with loop effects resulting in an LSP mass of $\sim 90$ GeV. In the non-split cases we choose $m_{NLSP}-m_{LSP} = 200 $ GeV so that the NLSP decays to the LSP + on-shell SM bosons.\end{comment}

\section{Squark-Gluino Associated Production}\label{sqgo}

In this section we discuss squark-gluino associated production. As this process only involves $\alpha_{s}$, it can be important at a $\sqrt{s} = 100$ TeV p-p collider even if $m_{\tilde{q}} + m_{\tilde{g}} \gtrsim 35$ TeV. If a heavy squark of order tens of TeV is produced in association with a gluino of mass $\lesssim 10$ TeV, the leading jet from the squark decay will be very hard, $p_T \sim m_{\tilde{q}}/2$. Furthermore the neutralino resulting from the decay chain $\tilde{q} \rightarrow q \tilde{g} \rightarrow 3\, q \chi^0$ will be very boosted, resulting in large $\slashed{\it{E}}_T$. These kinematic features result in a striking collider signature with very low SM background. 

We explore the reach in squark-gluino production at a $\sqrt{s} = 100$ TeV p-p collider for the two types of squark-gluino spectra listed in Table \ref{SimpModels.TAB}. For simplicity we assume the LSP is a Bino, and all other neutralinos/charginos are decoupled. Relaxing this assumption allows squark decays to intermediate neutralinos/charginos, resulting in additional final state objects which can be used for background discrimination. 

\begin{figure}[h]
\includegraphics[width=0.43\textwidth]{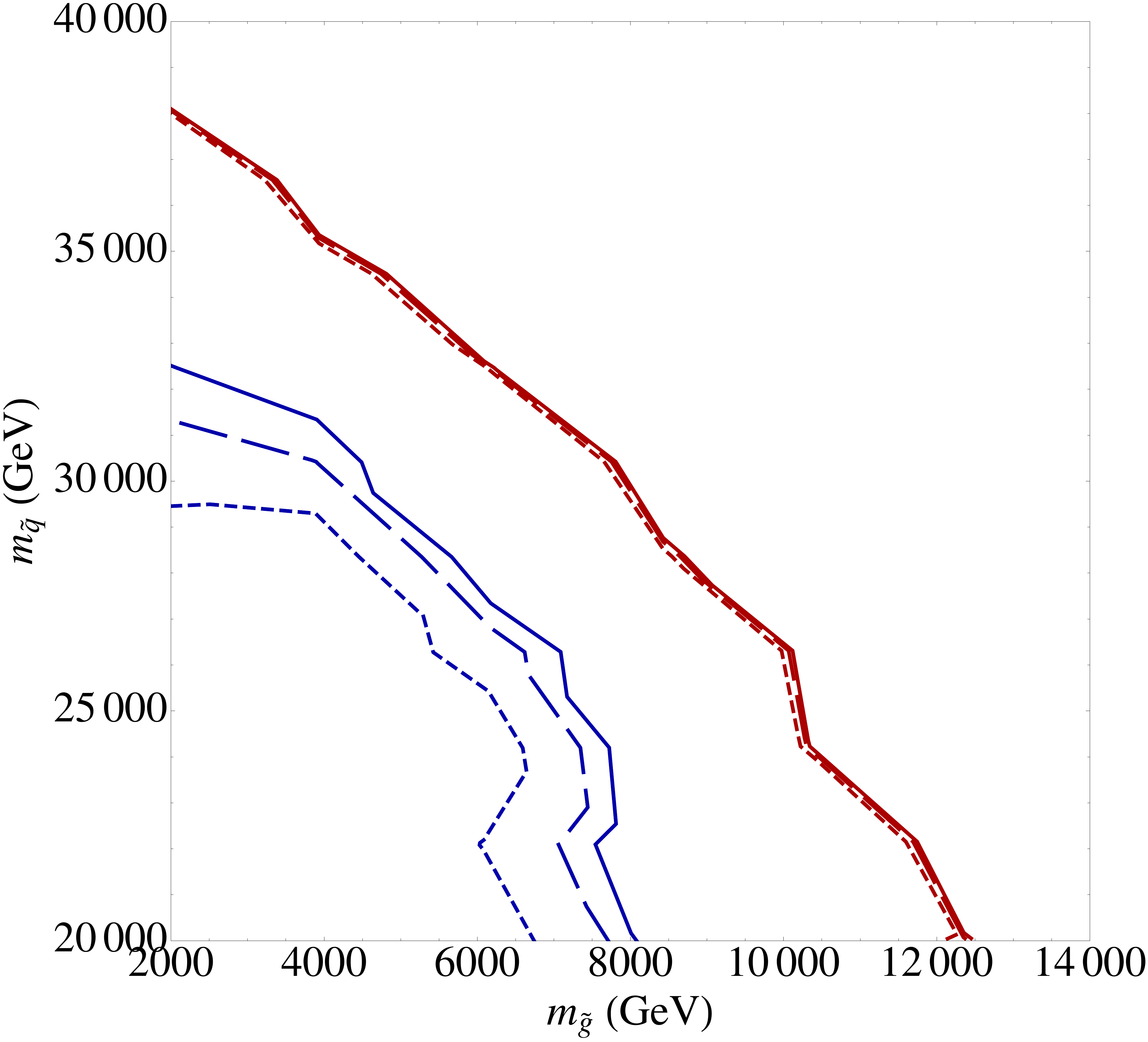}
\caption{Experimental reach for squark-gluino associated production at a 100 TeV proton collider with 3 ab$^{-1}$ integrated luminosity, for spectra with a $\sim 100$ GeV LSP mass. The solid, long dashed and short dashed lines are for and  $5,~10,~15$\% systematic uncertainty for the signal respectively. Blue lines indicate 5$\sigma$ discovery reach and red lines indicate 95\% exclusion limits. We assume 20\% systematic uncertainty in the background.}
\label{NonComp.FIG}
\end{figure}

\begin{figure}[h]
\includegraphics[width=0.43\textwidth]{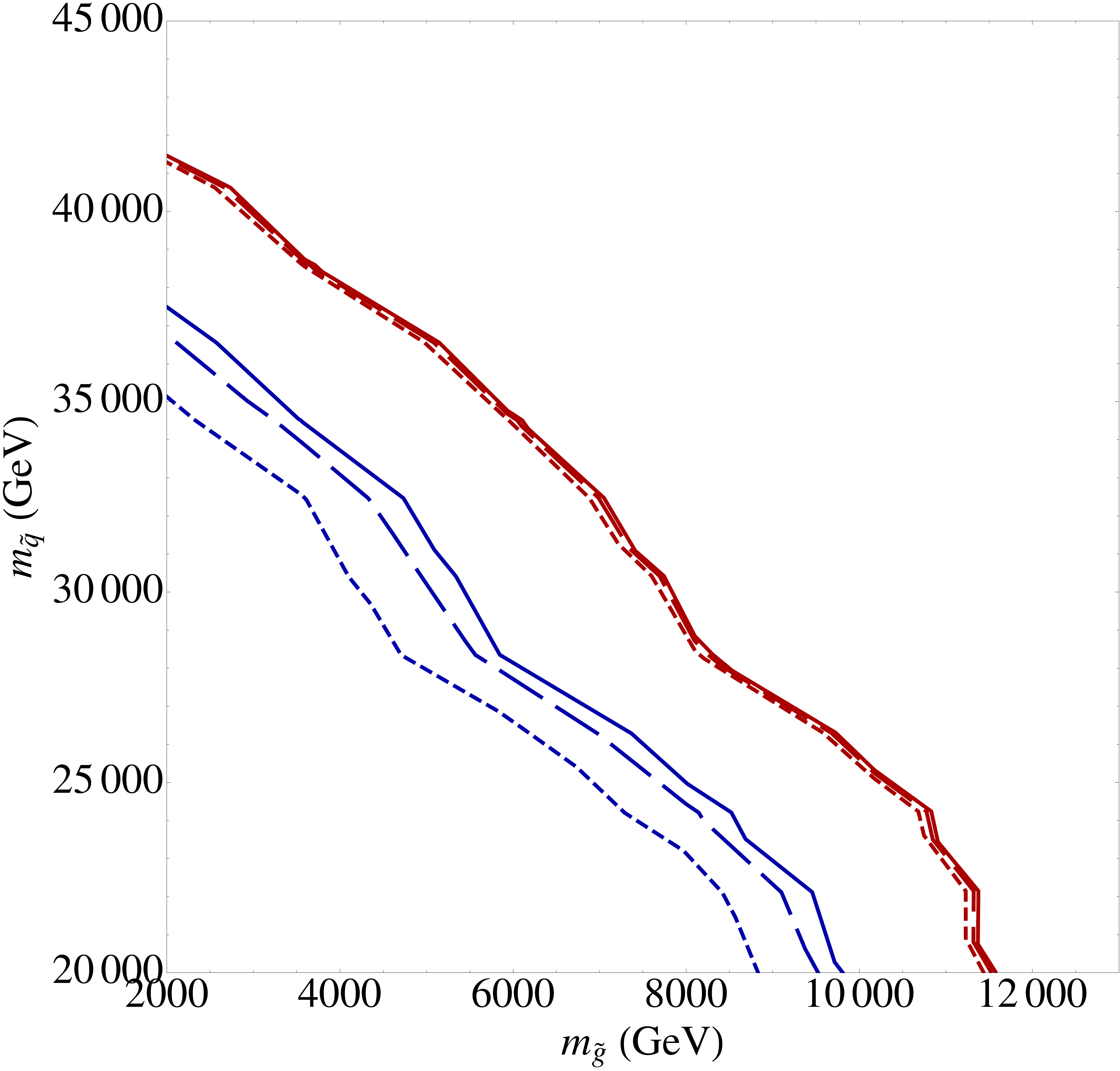}
\caption{Experimental reach for squark-gluino associated production at a 100 TeV proton collider with 3 ab$^{-1}$ integrated luminosity for spectra with $m_{\tilde{g}} - m_{\chi_1^0} = 15$ GeV. The different lines follow the conventions of Fig. \ref{NonComp.FIG}. We assume 20\% systematic uncertainty in the background.}
\label{Comp.FIG}
\end{figure}

For both non-compressed and compressed spectra, we impose the following baseline cuts:\begin{equation*} H_{T} > 10 \,\,\mathrm{TeV},\,\,\,\, \slashed{\it{E}}_T /\sqrt{H_T} > 20 \,\, \mathrm{GeV}^{1/2}\end{equation*} while for the non-compressed spectra we impose the additional cut:\begin{equation*}
8 \,\,\, \mathrm{jets \,\,\,with}\,\,\, p_T > 50\,\, (150) \,\, \mathrm{GeV}
\end{equation*} The softer cut is optimized for heavier squarks and lighter gluinos, while the harder cut is optimized for lighter squarks and heavier gluinos. Upon imposing these baseline cuts, we then scan over leading jet $p_T$ and $\slashed{\it{E}}_T$ cuts in order to maximize significance $\sigma$ as defined in (\ref{sig}).  We have verified that the optimal cuts render any ``background" from gluino pair production subdominant to the SM background.

The results of this analysis are depicted in Figs. \ref{NonComp.FIG} and \ref{Comp.FIG}, which show the reach of a $\sqrt{s} = 100$ TeV proton collider with 3 ab$^{-1}$ of integrated luminosity. The solid, long dashed and short dashed lines correspond respectively to systematic uncertainties of $5, ~10$ and $15$\% for the signal normalization, while the background systematic uncertainty is fixed to $20 \%$. The projected reach is fairly insensitive to background systematic uncertainties, as the number of background events is quite low due to the hard leading jet $p_T$ and $\slashed{\it{E}}_T$ cuts.

As is evident from Figs. \ref{NonComp.FIG} and \ref{Comp.FIG}, a $\sqrt{s} = 100$ TeV collider with 3 ab$^{-1}$ integrated luminosity can begin probing much of the ``mini-split" parameter space for sufficiently low gluino masses. Final states in the compressed spectra yield more $\slashed{\it{E}}_T$ compared to the non-compressed spectra, resulting in the greater reach depicted in Figure \ref{Comp.FIG}. Notably, with 3 ab$^{-1}$ integrated luminosity the entire neutralino-gluino coannihilation region (whose upper endpoint lies at $m_{\tilde{g}} \approx m_{\tilde{\chi}} \approx 8$ TeV \cite{Ellis:2015vaa}) can be excluded if the squark masses are $\lesssim 28$ TeV.

It is worthwhile to compare Figs. \ref{NonComp.FIG} and \ref{Comp.FIG} to projected reaches for gluino pair production. Our results for non-compressed spectra have some overlap with \cite{Cohen:2013xda}\footnote{A search optimizing over $H_{T}$ cuts as opposed to leading jet $p_T$ cuts was done in \cite{Cohen:2013xda}. For the spectra in Fig. \ref{NonComp.FIG}, the $H_T$ cut based analysis has a 3-5 TeV weaker reach in $m_{\tilde{q}} + m_{\tilde{g}}$ with respect to squark-gluino associated production.}, which considered both pair production and associated production in similar spectra with squark masses $\lesssim 24$ TeV. The results of \cite{Cohen:2013xda} indicate that gluino pair production will likely be the discovery channel for colored superpartners for the spectra in Fig. \ref{NonComp.FIG} provided $m_{\tilde{g}} \lesssim 14$ TeV. On the other hand, if the gluino and the LSP are nearly degenerate, searches for gluino pair production rapidly lose sensitivity \cite{Cohen:2013xda}. Thus if the gluino and the LSP are nearly degenerate as in the gluino-neutralino coannihilation scenario, squark-gluino associated production would be a potential discovery channel for colored superpartners.

\section{Squark-Wino and Squark-Bino Associated Production}\label{sqew}

In this section we discuss squark-Wino and squark-Bino associated production. These channels are particularly important if squark-gluino associated production is inaccessible due to a sufficiently heavy gluino mass\footnote{In the MSSM, a gluino which is hierarchically heavier than the squarks requires fine-tuning of the soft masses. This can be avoided however in a model with Dirac gluinos \cite{Fayet:1978qc,Fox:2002bu}.}. The event topology is qualitatively similar to squark-gluino production, as the squark will decay to a boosted jet and boosted Wino/Bino while the associated Wino/Bino is produced at relatively low $p_T$. However as noted in Section \ref{general}, associated squark-Wino/Bino production probes significantly lighter squark masses than squark-gluino production. Consequently, multi-TeV leading jet $p_T$ and $\slashed{\it{E}}_T$ cuts are not as effective for background discrimination in squark-Wino/Bino production. Instead, we find that hard $\slashed{\it{E}}_T /\sqrt{H_T}$ cuts are quite effective at reducing the $t \overline{t} + $ jets and vector boson + jets background without rejecting too many signal events.

In order to determine the projected reach for squark-Wino/Bino production at a $\sqrt{s} = 100$ TeV pp collider with 3 ab$^{-1}$ integrated luminosity,  we impose the following baseline cuts:\begin{equation*} p_T(j_1) > 2 \,\,\mathrm{TeV},\,\,\,\, \slashed{\it{E}}_T > 3 \,\, \mathrm{TeV}, \,\,\, \Delta \phi (j,\slashed{\it{E}}_T) > 0.5 \end{equation*} where the $\Delta \phi$ cut is imposed only on the two leading jets. We then scan over $\slashed{\it{E}}_T /\sqrt{H_T}$ cuts for each spectrum to maximize $\sigma$ as defined in (\ref{sig}). 

\begin{figure}[h]
\includegraphics[width=0.43\textwidth]{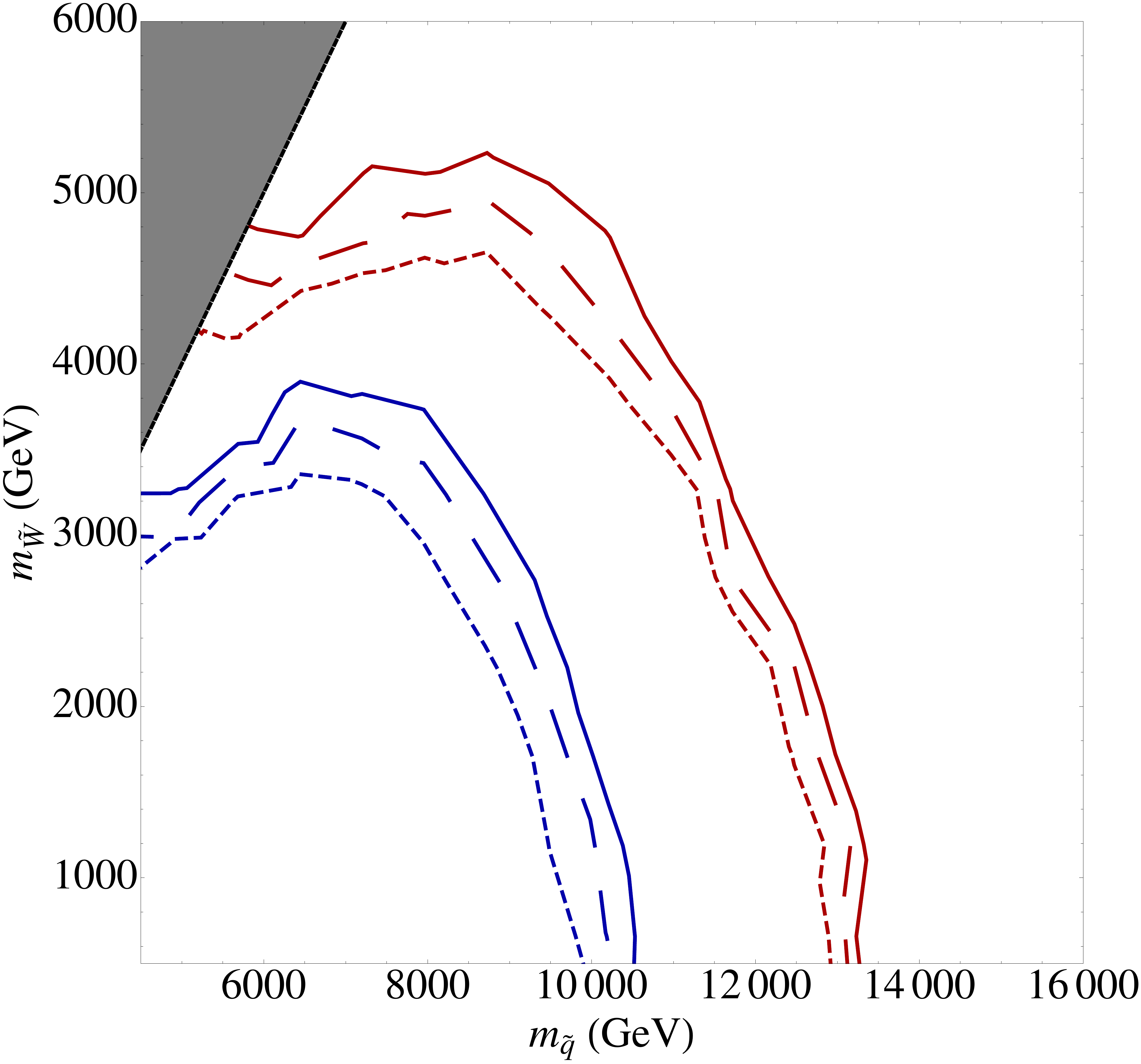}
\caption{Experimental reach for squark-Wino LSP associated production at a 100 TeV proton collider with 3 ab$^{-1}$ integrated luminosity. The solid, long dashed and short dashed lines are for  $1,~2,~3$\% systematic uncertainty for the background respectively. Blue lines indicate 5$\sigma$ discovery reach and red lines indicate 95\% exclusion limits. We do not consider the grey shaded region ($m_{\tilde{q}}-m_{\tilde{W}} < 1$ TeV) for reasons given in the text. We assume 10\% systematic uncertainty for the signal.}
\label{SqWino.FIG}
\end{figure}

\begin{figure}[h]
\includegraphics[width=0.43\textwidth]{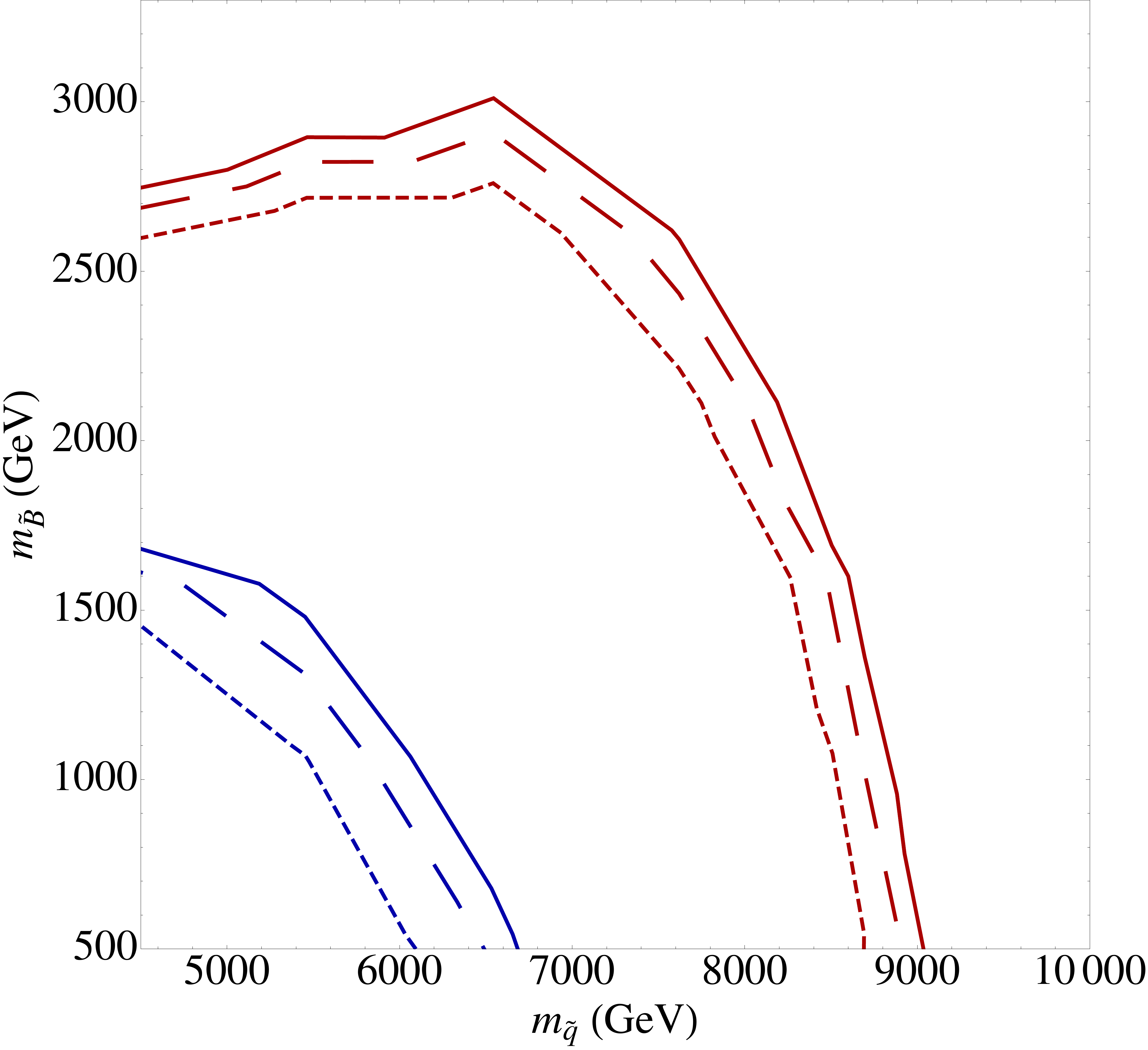}
\caption{Experimental reach for squark-Bino LSP associated production at a 100 TeV proton collider with 3 ab$^{-1}$ integrated luminosity.  The solid, long dashed and short dashed lines are for and  $0.5,~1,~1.5$\% systematic uncertainty for the background respectively. Blue lines indicate 5$\sigma$ discovery reach and red lines indicate 95\% exclusion limits. We do not consider the region ($m_{\tilde{q}}-m_{\tilde{B}} < 1$ TeV) for reasons given in the text. We assume 10\% systematic uncertainty in the signal.}
\label{SqBino.FIG}
\end{figure}

\begin{figure}[h]
\includegraphics[width=0.43\textwidth]{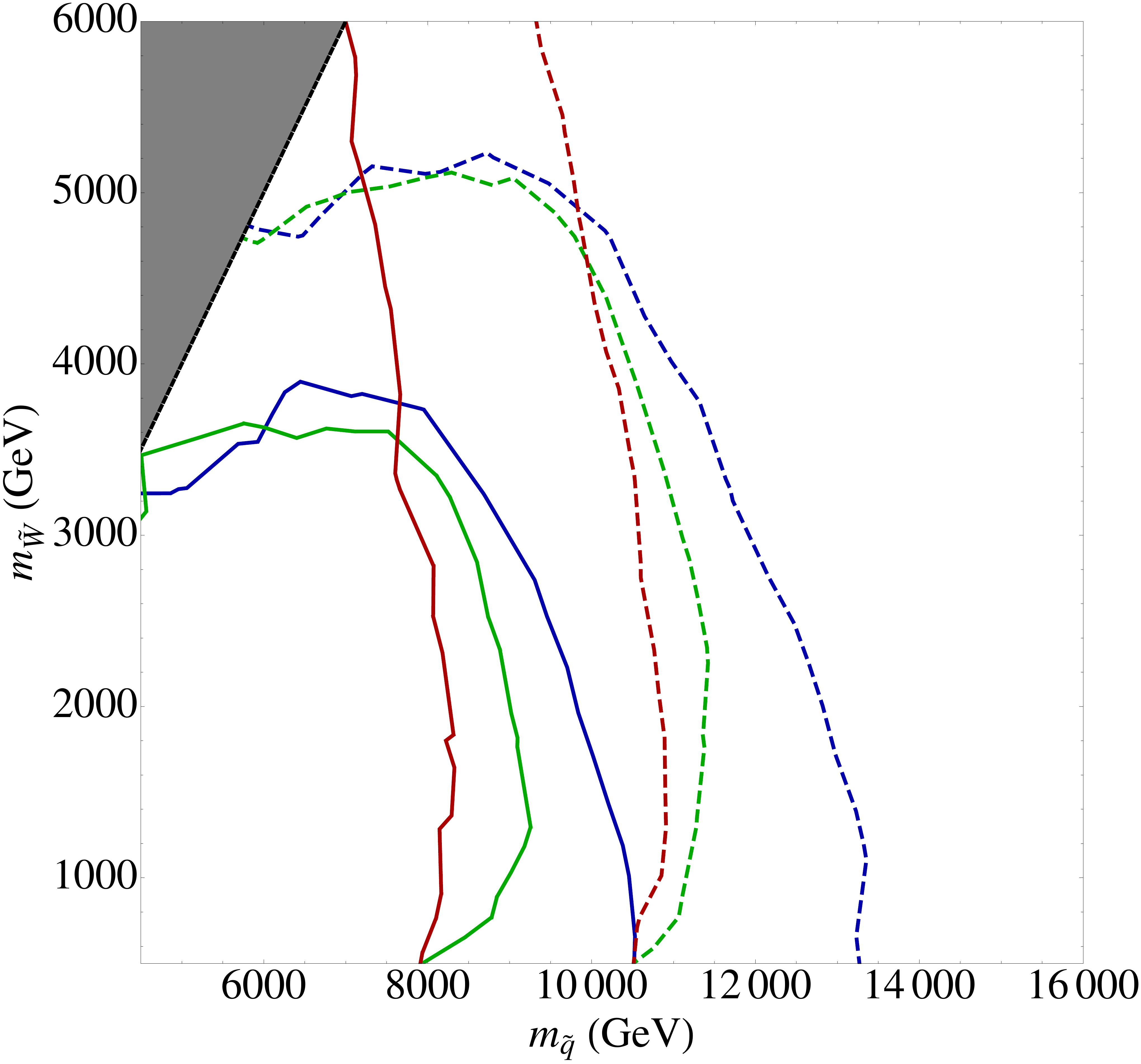}
\caption{Experimental reach for squark-Wino associated production at a 100 TeV proton collider with 3 ab$^{-1}$ integrated luminosity. Solid lines indicate 5$\sigma$ discovery reach, and dotted lines indicate 95\% exclusion limits. Blue curves correspond to a Wino LSP, while the green (red) curves correspond to a Wino NLSP with  $M_{\mathrm{NLSP}} - M_{\mathrm{LSP}}=200$ GeV ($M_{LSP} \sim 100$ GeV). The results are applicable for both Bino- and Higgsino-like LSP.  We do not consider the grey shaded region ($m_{\tilde{q}}-m_{\tilde{W}} < 1$ TeV) for reasons given in the text. We assume 1\% systematic uncertainty in the background and 10\% in the signal.}
\label{SqWinoNLSP.FIG}
\end{figure}

Our focus is on spectra listed in Table \ref{SimpModels.TAB} where at most one of the gaugino/Higgsino mass parameters $M_1,\, M_2, \, \mu$ are $\lesssim 1$ TeV, such that the gauge eigenstates are approximately aligned with the mass eigenstates in the neutralino/chargino sectors. We omit the ``compressed" region $m_{\tilde{q}} - m_{\tilde{\chi}} < 1 $ TeV, as in this region the event topology of associated squark-Wino/Bino production is similar to squark pair production, only with a substantially smaller cross section. Assuming a systematic uncertainty of $10 \%$ for the signal normalization, the results of the above analysis for the various spectra in Table \ref{SimpModels.TAB} are depicted in Figures \ref{SqWino.FIG}-\ref{SqWinoNLSP.FIG}. 

Figure \ref{SqWino.FIG} shows the reach for squark-Wino production with a pure Wino LSP; the solid, short-dashed, long-dashed lines correspond to background uncertainties of 1\%, 2\% and 3 \%. In Figure \ref{SqBino.FIG} we show the reach for squark-Bino production with a pure Bino LSP. The solid, short-dashed, long-dashed lines correspond to background systematic uncertainties of 0.5\%, 1\% and 1.5\%. Compared to squark-Wino production, the reach for squark-Bino associated production is quite sensitive to background uncertainties. This is because the 5$\sigma$ contours for squark-Bino production correspond to significantly lower masses due to the smaller production cross-section, resulting in lower optimal $\slashed{\it{E}}_T /\sqrt{H_T}$ cuts and thus larger backgrounds. 

In Figure \ref{SqWinoNLSP.FIG} we show the reach of the $\slashed{\it{E}}_T /\sqrt{H_T}$ based monojet analysis for squark-Wino production with a Wino NLSP, with background uncertainties fixed to be 1\%. The green lines correspond to $M_{NLSP} - M_{LSP} = 200$ GeV, while the red lines correspond to $M_{LSP} = 100$ GeV. For comparison, the blue lines show the reach for squark-Wino production when the Wino is the LSP. Away from the $m_{\tilde{q}} \sim m_{\tilde{W}}$ region the sensitivity is lower for a Wino NLSP, as $\slashed{\it{E}}_T$ is being traded for $W, Z$ and higgs bosons arising from the NLSP $\rightarrow$ LSP decay. Note that the analysis considered here does not exploit the additional SM bosons present in the Wino NLSP scenario. Thus the reach for the Wino NLSP scenario depicted in Figure \ref{SqWinoNLSP.FIG} applies regardless of whether the LSP is Bino-like or Higgsino-like. Exploiting the additional SM bosons could extend the reach for the Wino NLSP scenario, so the result presented here is a conservative estimate.

We close this section by comparing the results of Figures \ref{SqWino.FIG}-\ref{SqWinoNLSP.FIG} to studies of pair production at $\sqrt{s} = 100$ TeV. Given 3 ab$^{-1}$ integrated luminosity, squark pair production can discover squark masses up to $2.5$ TeV \cite{Cohen:2013xda} (assuming a conservative 20 $\%$ background systematic uncertainty). In the pure Wino case, searches in VBF channels can discover Winos up to 1.1 TeV \cite{Berlin:2015aba}. Disappearing tracks can also provide a collider probe of pure Wino LSP pair production. Extrapolating the disappearing tracks background from the 8 TeV ATLAS study \cite{Aad:2013yna}, the projected reach is 2-3 TeV for pure Winos \cite{Low:2014cba}. However, the data-driven disappearing-track background at 100 TeV is difficult to estimate, making this projected reach less reliable than the reach in the VBF channel or the reach depicted in Figure \ref{SqWino.FIG}. Finally, pair production of Wino NLSPs has been considered in \cite{Acharya:2014pua, Gori:2014oua}. Assuming no systematic uncertainties, for a Higgsino LSP the projected discovery reach is 2.3 TeV, while for a Bino LSP the reach is 1-3 TeV depending on the NLSP $\rightarrow Z$ LSP branching ratio. Comparing these reaches to Figures \ref{SqWino.FIG}-\ref{SqWinoNLSP.FIG}, we see that squark-Wino/Bino associated production can provide a SUSY discovery mode provided the squark is not too much heavier than the Wino/Bino.

\section{Summary}\label{conclusion}
We have examined in this paper the kinematic reach for squark-gaugino associated production at a 100 TeV proton proton collider. In models where squark pair production is kinematically inaccessible at a 100 TeV collider, squark-gaugino associated production may be the discovery mode for SUSY in a large portion of parameter space. 

We have considered the various simplified models listed in Table \ref{SimpModels.TAB}. For squark-gluino production with $\mathcal{O}$(TeV) gluinos, the discovery reach for first-generation squarks can be up to 37 TeV for compressed spectra (small gluino-LSP mass splitting), and up to 32 TeV for non-compressed spectra, subject to systematic uncertainties. For squark-Wino LSP production, we have shown that the discovery reach for the Wino is almost 4 TeV for squarks of $\sim7$ TeV, subject to systematic uncertainties. For squark-Wino NLSP production we have analysed two scenarios: one where the NLSP-LSP mass difference is 200 GeV, and one where the LSP mass is $\sim 100$ GeV. In the first scenario, the Wino discovery reach is about 3.5 TeV for squarks of $\sim 7$ TeV. In the second scenario, the Wino reach extends up to 6 TeV. Our results in the Wino-NLSP scenario are insensitive to the nature of the LSP. For $\lesssim 9$ TeV squark masses, squark-Wino associated production marks a significant increase in the Wino reach compared to pair production channels. We also consider squark-Bino associated production, and find that the kinematic reach for the Bino is up to 1.7 TeV for squarks of mass $\sim5$ TeV, subject to systematic uncertainties. 

The results presented here raise the exciting prospect of directly probing a region of parameter space that so far has been the exclusive domain of indirect searches through low-energy FCNC observables. The squark-gaugino associated production channels studied here, coupled with studies of supersymmetry at 100 TeV colliders already undertaken \cite{Cohen:2013xda, Jung:2013zya, Low:2014cba, Cohen:2014hxa, Ellis:2014kla, Acharya:2014pua, Gori:2014oua, Bramante:2014tba, diCortona:2014yua, Berlin:2015aba, Beauchesne:2015jra}, provide a strong physics case for the construction of such a collider.

\section*{Acknowledgements}

We would like to thank T. Cohen, J. Ellis, T. Han, G. Kane, M. Low, K. Olive, N. Orlofsky, A. Pierce and L.T. Wang for useful discussions. The work of S.A.R.E. and B.Z. was supported in part by DOE grant DE-SC0007859.

\appendix
\section{Event generation}
\label{GenDetails.APP}

Signal events were generated using \verb MADGRAPH5  \cite{Alwall:2014hca}, with showering and hadronization implemented via \verb PYTHIA6.4  \cite{Sjostrand:2006za}. We do not perform MLM for the signal events. We have validated this approximation by performing MLM with 2 additional jets for a number of benchmark spectra. We use the simulated Snowmass backgrounds \cite{Avetisyan:2013onh}, proccessed with \verb Delphes3.1.2  \cite{deFavereau:2013fsa} supplemented by the Snowmass detector card \cite{Anderson:2013kxz} for a $\sqrt{s} = 100$ TeV hadron collider. Production cross sections for squark-gluino associated production are computed at NLO using \verb PROSPINO2  \cite{Beenakker:1996ed}. For squark-Wino/Bino production we use the LO result computed by \verb MADGRAPH5. Event analysis is performed with \verb MadAnalysis 5 \cite{Conte:2012fm}. We expect our kinematic cuts to effectively remove any contamination from QCD backgrounds and pileup effects, so we neglect both of these in our analysis. Note that for squark-gluino associated production in the $m_{\tilde{q}} \gg m_{\tilde{g}}$ region, the dijet background may not be negligible for non-compressed spectra. For these spectra, jet substructure techniques can help distinguish signal events from the QCD background \cite{Fan:2011jc}.

\end{document}